\documentclass[article]{revtex4}
\usepackage{graphicx}
\usepackage{amssymb}
\usepackage{subfig}
\usepackage{color}

\begin{document}

\title{Rydberg atoms for radio-frequency communications and sensing: atomic receivers for pulsed RF field and phase detection}
\author{D. A. Anderson (email: dave@rydbergtechnologies.com)}
\author{R.E.~Sapiro}
\author{G. Raithel}
\affiliation{Rydberg Technologies Inc., Ann Arbor, MI 48103 USA}
\date{\today}
\maketitle

\section{Introduction}
\label{sec:intro}

The emergence of atomic sensor technologies is driving a paradigm shift in modern sensing and measurement by exploiting quantum phenomena to realize fundamentally new detection capabilities unmatched by their classical counterparts~\cite{MacFarlane.2003}.  Atomic sensing of radio-frequency (RF) electric fields using Rydberg electromagnetically-induced transparency (EIT) in atomic vapors has been the subject of growing scientific interest~\cite{Sedlacek.2012,Holloway2.2014,Anderson.2016}.  This has been motivated in part by a drive at National Metrology Institutes to replace century-old antennas as RF standards with absolute (atomic) standards for RF electric fields~\cite{Holloway2.2014}, and has recently been established as a novel quantum technology platform with broad capabilities~\cite{Anderson2.2017, Anderson.2019} that has matured into commercial RF detection and measurement instrumentation~\cite{RFMS.2019,RydbergTech}.  A notable advance in atomic RF devices and measurement tools is the recent realization of the first Rydberg RF field probe (RFP) and measurement system (RFMS) for self-calibrated SI-traceable broadband RF measurement and imaging of continuous, pulsed, or modulated fields.  Relevant developments include the realization of compact atomic sensing elements capable of broadband RF electric-field measurement from MHz to $>$100~GHz~\cite{AndersonGSMM.2018}, fiber-coupled atomic vapor-cell RF field probes~\cite{SimonsProbe.2018,Anderson.2019}, the demonstration of ultra-wide dynamic field ranges spanning sub-10~mV/m up to $>$10~kV/m (dynamic range $>$120~dB)~\cite{Kuebler.2019,Paradis.2019}, and all-optical circuit-free RF sensors for EMP/EMI-tolerant detection and operational integrity in high-intensity RF environments~\cite{Anderson.2017}.  Hybrid atomic RF technology that combines atom-based optical sensing with traditional RF circuitry and resonators has also been developed realizing hybrid sensors with augmented performance capabilities such as resonator-enhanced ultra-high sensitivity polarization-selective RF detectors~\cite{Anderson2.2018}, waveguide-embedded atomic RF E-field measurement for SI-traceable RF power standards~\cite{HollowayPower.2018}, and atom-mediated optical RF-power/voltage transducers and receivers~\cite{SapiroDAMOP.2019}.  Recently, Rydberg atom-based field sensing has also been adapted to modulated RF field detection promising new possibilities in RF communications, with demonstrations including a Rydberg-atom transmission system for digital communication~\cite{Meyerpub.2018}, atom radio-over-fiber~\cite{Debpub.2018}, and ``Atomic Radio"~\cite{GetReadyForAtomicRadio} using a multi-band atomic AM and FM radio receiver based on direct atom-mediated RF-to-optical conversion of baseband signals picked up from modulated RF carriers~\cite{RydbergRadio.2018}. \\


In this article we describe the basic principles of the atomic RF sensing method and present the development of atomic pulsed RF detection and RF phase sensing establishing capabilities pertinent to applications in communications and sensing.  To date advances in Rydberg atom-based RF field sensors have been rooted in a method in which the fundamental physical quantity being detected and measured is the electric field amplitude, $E$, of the incident RF electromagnetic wave. Sections~\ref{sec:field} and~\ref{sec:com} are focused on using atom-based $E$-field measurement for RF field-sensing and communications applications. With established phase-sensitive technologies, such as synthetic aperture radar (SAR) as well as emerging trends in phased-array antennas in 5G, a method is desired that allows robust, optical retrieval of the RF phase using an enhanced atom-based field sensor.  Section~\ref{sec:phase} is focused on a fundamentally new atomic RF sensor and measurement method for the phase of the RF electromagnetic wave that affords all the performance advantages exhibited by the atomic sensor~\cite{Anderson.2019}.  The presented phase-sensitive RF field detection capability opens atomic RF sensor technology to a wide array of application areas including phase-modulated signal communication systems, radar, and field amplitude and phase mapping for near-field/far-field antenna characterizations.




\section{Atomic-physics and field/phase-sensing background}

\label{sec:background}

Our atom-based field sensors use Rydberg atoms as an RF-receiver medium.
Classically, a Rydberg state is a state of an atom in which a valence electron resides in an orbit far from the atomic core.  The weakly-bound, quasi-free electron of a Rydberg atom affords the atom a unique set of physical properties including a high sensitivity to external electric and magnetic fields. The atomic-physics principles of one- and two-electron systems are described by Bethe and Salpeter~\cite{Bethe.1957}. Rydberg atoms of alkali, earth alkali and a variety of other species fall within this class of atomic systems.  Several textbooks that are specifically focused on the physics of Rydberg atoms include the works by Gallagher~\cite{GallagherBook} and Stebbings and Dunning~\cite{DunningBook}. For the present purpose Rydberg atoms may be viewed as quantum oscillators that are fairly easy to prepare via laser excitation, and that are perfectly frequency-matched to a selection of incident RF frequencies. This is because the orbital frequencies of the Rydberg valence electron can be tuned into resonance with RF radiation. The set of highly responsive frequencies is different for every Rydberg state. Since there is a wide variety of different Rydberg states that are accessible by tuning the Rydberg-atom excitation lasers, Rydberg atoms offer broadband RF coverage from the MHz into the THz regime.\\

A single Rydberg-atom receiver consists of a valence electron of a single atom that has been laser-excited into a Rydberg state, whose orbital frequency allows a (near-)resonant, RF-driven transition into another Rydberg state. The frequency match affords a combination of very small receiver size, and high electric-field sensitivity. A single receiver Rydberg atom has a size on the order of a $\mu$m, while an atomic ensemble that is large enough for the construction of a technically viable and robust receiver instrument can range between hundreds of $\mu$m and a few cm in size. The response of the atomic ensemble to an incident RF field amounts to quantum-mechanical energy level splittings and level shifts that are observed by the means of EIT laser beams, which present an all-optical, robust tool to measure the atomic response, and to thereby determine the RF field. As the measurement is based on invariable atomic properties that are well known, this method of RF field determination is atom-based and intrinsically calibration-free. In Sections~\ref{sec:field} and~\ref{sec:com} we employ a Rydberg-atom field sensor to measure RF field amplitudes and to receive modulated RF signals.\\

To achieve phase sensitivity in an atom-based Rydberg receiver, we employ elements of holographic phase-sensing methodologies. The phase of the signal wave, $\phi$, is defined relative to the phase of a reference oscillator or reference wave, $\phi_{ref}$. To enable phase measurement of the signal wave, the signal electromagnetic field has to be brought into an interferometric relationship with the reference field. In practice, the phase reference is
often mediated via a reference wave that is physically superimposed with the signal wave on top of a detector that measures field amplitude. Due to the superposition principle, which is common to all wave phenomena that follow the (linear) wave equation, the phase difference, $\phi - \phi_{ref} - \phi_{ofs}$, is obtained from an interference measurement. In its most basic implementation, the net signal is given by the sum of two sine waves with the same frequency, $A \sin(\omega t + \phi) + A_{ref} \sin (\omega t + \phi_{ref} + \phi_{ofs})$, with amplitudes $A$ and a controllable offset phase $\phi_{ofs}$ that is used to tune the interference pattern from constructive to destructive, and to thereby find a value for $\phi - \phi_{ref}$. A measurement of the net wave amplitude versus $\phi_{ofs}$ yields the phase difference between the wave to be tested and the reference wave, $\phi - \phi_{ref}$. This usually sums up the task of phase measurement. The principle of differential-phase measurement by the means of superposition of object and reference waves is widely used in holography. There, phase- and amplitude-sensitive recordings of interference patterns of signal- and reference waves on a planar recording medium with sub-wavelength spatial resolution allow accurate, three-dimensional reconstruction of the signal wave field. This holography concept can be translated from the optical into the RF domain. In Section~\ref{sec:phase} we describe a method of atom-based RF phase detection, measurement and enhanced receiving that we have recently devised. The method is not limited to signal and reference waves of the same RF frequency. Reference waves that are offset in frequency enable heterodyne and superheterodyne signal amplitude and phase detection.

\section{Atomic RF electric field sensing}
\label{sec:field}

\begin{figure}[h]
\centerline{\includegraphics[width=\columnwidth]{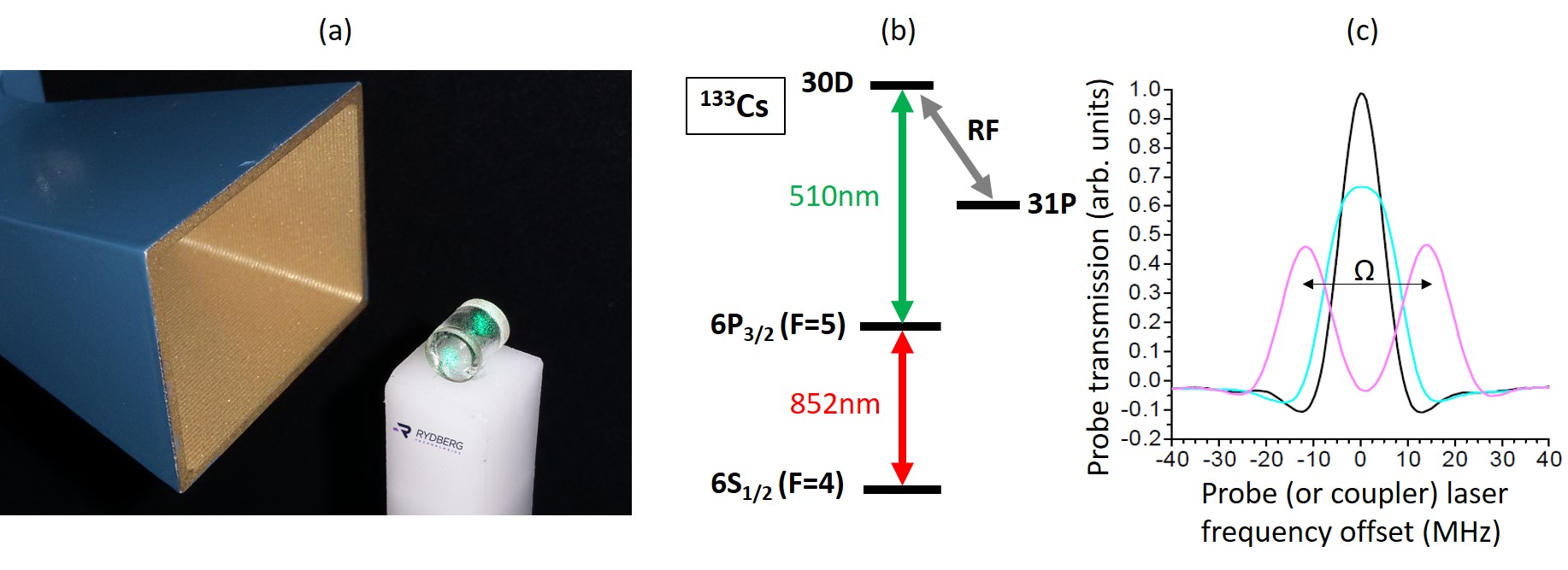}}
\caption{(a) An atomic vapor-cell RF sensing element in front of a K$_{a}$-band horn antenna. Photo courtesy of Rydberg Technologies Inc.~\cite{RydbergTech}, (b) atomic energy-level diagram for RF sensing and measurement using a two-photon Rydberg EIT scheme in cesium, (c) Rydberg EIT signal readout from Rydberg state without RF (black), with on-resonant weak RF field (teal) and moderate RF field exhibiting Autler-Townes splitting (magenta).}
\label{fig1}
\end{figure}

Atomic RF receiver technology employs EIT as a quantum-optical readout of Rydberg states of atoms in a vapor~\cite{Mohapatra.2007,Sedlacek.2012,Holloway2.2014,Anderson.2016}. Figure~\ref{fig1}(a) shows a picture of a miniature atomic vapor cell sensing element containing a pure cesium gas next to a standard K$_a$-band horn antenna.  To sense and measure parameters of an incident RF field, optical beams are passed through the vapor cell to interrogate field-sensitive Rydberg states of the atoms exposed to the RF field.  Detected changes in the transmission of an optical probing beam through the atomic vapor provide a direct RF-to-optical readout and information on the incident RF signal field.  Under typical
operating conditions, the atomic vapor has an optical density for the EIT probe laser beam propagated through the cell that is sufficiently high to obtain a robust EIT signal with high signal to noise, as required for RF field detection.  Further, the atomic vapor in the cell is dilute enough so that interactions of the Rydberg atoms can  be neglected. Therefore, the spectroscopic response of the medium to the fields can be modeled based on a quantum-mechanical picture of a single, isolated atom.\\

Figure~\ref{fig1}(b) shows an atomic energy-level diagram illustrating a two-photon Rydberg EIT readout scheme for a cesium vapor.  In this basic scheme, two optical laser fields couple atomic states to a high-lying Rydberg state (30D in Fig.~\ref{fig1}(b)), with a weak optical probe beam resonant with the first atomic transition between ground and intermediate states, and a relatively stronger optical coupler beam tuned into resonance with a second atomic transition between intermediate and Rydberg state.  When the coupler laser-frequency is in resonance with the Rydberg state, an EIT window opens for the probe beam through the vapor~\cite{Fleischhauer.2005,Berman}. Owing to the sensitivity of the atomic Rydberg levels to the RF electric field, the Rydberg EIT signal provides an optical readout for the RF field. An example Rydberg EIT resonance is shown in Figure~\ref{fig1}(c) (black curve).  In the presence of a moderate RF field at a frequency near-resonant with an allowed transition between the optically excited Rydberg level and a second Rydberg level of the atom, the EIT-detected atomic Rydberg line splits into a pair of Autler-Townes (AT) lines whose splitting is proportional to the RF electric-field amplitude (Figure~\ref{fig1}(c) (magneta curve)). In this linear AC Stark effect regime, the E-field is given by~\cite{Berman}

\begin{equation}
E = \hbar \Omega /d,
\end{equation}

\noindent where $\Omega$ is the Rabi frequency of the RF-coupled atomic Rydberg transition (near-identical to the AT splitting measured optically in units $2\pi \times $Hz), $d$ is the electric dipole moment of the Rydberg transition in units Cm, and $\hbar=6.62606 \times 10^{-34}$~Js$/(2\pi)$ is Planck's constant.\\

From Eq.~1 one obtains a direct, optical measurement of the electric field amplitude $E$ of the RF wave in absolute (atomic) units traceable to fundamental constants. Generally, for low RF field levels, the sensitivity of the atomic receivers is dictated by (1) the electric dipole moment $d$ of the Rydberg-Rydberg transition resonant with the incident RF field and (2) the spectroscopic EIT linewidth in the optical readout which determines the achievable resolution for measuring $\Omega$.  For RF-field frequencies in the range of 100~MHz to 500 GHz, resonant dipole moments in alkali atoms typically range from $10^{2}$ to $10^{5}$~ea$_0$, where $e$ is the elementary charge and a$_0$ is the Bohr radius, with the principal quantum number $n$ ranging from about 10 to 300, while Rydberg EIT linewidths are typically about 1~MHz or more.  Equation~1 provides a useful approach to RF E-field sensing and measurement with EIT in Rydberg atom vapors, but serves to a large extent as a didactic model because it is valid only within a relatively limited E-field range and for a discrete, albeit large, set of RF field frequencies near-resonant with Rydberg transitions, thereby rendering it impractical in many real-world E-field measurement scenarios.  This is addressed by a well-developed measurement method and approach using EIT and exploiting the full quantum response of the Rydberg atom interaction with RF fields that includes off-resonance AC Stark shift readouts~\cite{AndersonGSMM.2018}, enabling direct $E_{RMS}$ measurement of continuous-frequency RF field frequencies over tens of GHz with a $>$60~dB dynamic range. A full non-perturbative Floquet treatment allows measurement of the electric-field values and frequencies of even stronger (coherent) RF fields~\cite{Anderson.2016,Anderson2.2017}.\\

\section{Communications and modulated RF field sensing with atomic receivers}
\label{sec:com}

The adaptation of the Rydberg atom-based RF E-field sensing and measurement approach to the detection of modulated and time-varying RF fields promises to enable new capabilities in RF sensing and communications~\cite{Anderson.2019}.  Recent laboratory work has been performed demonstrating modulated RF E-field detection and baseband signal reception with Rydberg EIT in atomic vapors.   Highlights include a Rydberg atom-based transmission system for digital communications~\cite{Meyerpub.2018}, atom radio-over-fiber~\cite{Debpub.2018}, and a multi-band atomic AM and FM receiver for radio communications~\cite{RydbergRadio.2018} recently adapted to two-channel reception using two atomic species~\cite{HollowayRydbergRadio.2019}.  Atomic receivers for communications are a nascent technology prime for advanced development and adaptation to real-world systems.\\

The basic operating principle of an atomic RF receiver based on Rydberg EIT in vapor cells exploits the large differential dipole moments of Rydberg states of atoms.  With an RF carrier wave applied to the atomic sensing volume, the coupler-laser frequency is set to an operating point on one of the inflection points of the EIT spectral line (see, for example, Figs.~\ref{fig1}(b) and (c)).   As the incident modulated RF wave impinges on the atoms, the atoms respond synchronously to the time-varying RF electric field leading to a change in the probe light transmission through the vapor.  This realizes a direct Rydberg-atom-mediated optical pick-up and demodulation of the baseband-modulated RF carrier signal, where the demodulation occurs in the atomic vapor cell without need for any demodulation or signal-processing electronics required by traditional antenna-receiver technology.\\

\begin{figure}[h]
\centerline{\includegraphics[width=\columnwidth]{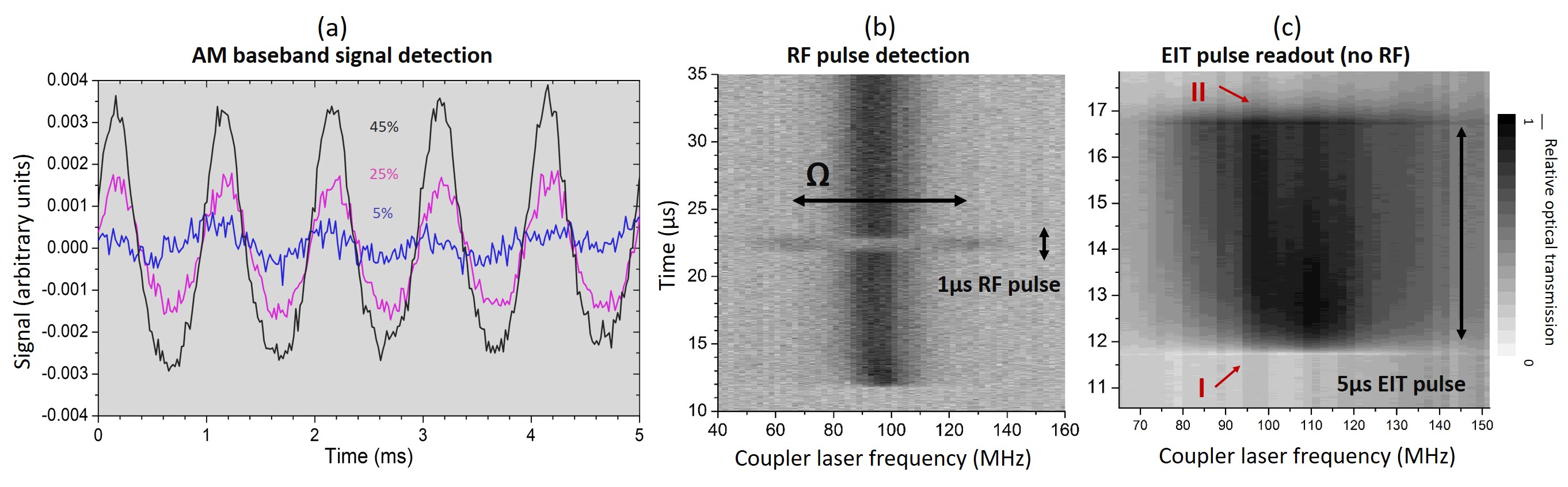}}
\caption{(a)Real-time optical readout from the atomic receiver for an AM 1~kHz baseband signal transmitted on a 37.406~GHz RF carrier resonantly driving the cesium 47S$_{1/2}$ to 47P$_{1/2}$ Rydberg transition. The received signals for three AM modulation depths of ±5\% (blue), ±25\% (purple), and ±45\% (black) are shown for the coupler laser-frequency operating point set to the field-free cesium 47S$_{1/2}$ Rydberg line~\cite{RydbergRadio.2018} (b) time-domain detection and measurement of a 36.2-GHz 1~$\mu$s-long RF field pulse using a rubidium Rydberg-based atomic detector as a function of time and coupler-laser frequency. The time evolution is along the y-axis.  The coupler laser beam is switched on at 11.7~$\mu$s and left on, and the RF pulse incident on the sensing element is switched on at 21.7~$\mu$s. The RF frequency is resonant with the rubidium 47S$_{1/2}$ to 47P$_{1/2}$ Rydberg transition (dipole moment $d=745$~ea$_0$) and produces a splitting (double-sided arrows) of the EIT line proportional to the pulsed RF field amplitude of about 5~V/m. (c) Relative EIT probe transmission for a 5~$\mu$s-long coupler laser square-pulse (probe laser on continuously) as a function of time and coupler-laser frequency near the rubidium 30D$_{5/2}$ Rydberg state.  No RF is applied.  The coupler pulse is on at 11.7~$\mu$s and off again at 16.7$\mu$s with an uncertainty $<$100~ns.  The probe transmission is in gray-scale; coupler-free absorption background is at a level of 0.236; with relative increasing transmission from white to black.} 
\label{fig2}
\end{figure}

For the general case of a transmitted AM signal and differential dipole moment $d$ of the target Rydberg states in the atomic receiver, a typical range in AM depth $\delta E/E$ is given by $\delta E/E \sim h \times \delta\Gamma/(E d)$, where $\delta\Gamma$ is the EIT linewidth.  Figure~\ref{fig2}(a) shows the real-time optical readout from an atomic rubidium vapor-cell receiver detecting and demodulating 1~kHz baseband signals transmitted in free space on an AM-modulated 37.4065~GHz RF carrier wave.  Received signals are shown for three different AM modulation depths of the carrier.  The modulation depths can typically range from several 10\% down to below 1\%, depending on exact operating conditions and receiver sensitivity requirements. In addition to being sensitive to changes in RF field amplitude, Rydberg states are also sensitive to changes in RF field frequency, allowing receiver pick-up and demodulation of FM RF carrier signals using a similar approach.  This basic approach has been implemented in the reception of both AM and FM radio communications on RF carrier waves over a wide range of carrier bands, with wide-band operation of a single atomic receiver demonstrated for carrier frequencies spanning more than four octaves, from C-band to Q-band.\\




In addition to radio and digital communications, pulse-modulated RF field detection and measurement with Rydberg atom receivers promises to expand atom-based RF technology for enhanced performance capabilities in application areas including high-intensity pulsed-RF measurement and electromagnetic testing, pulsed radar, surveillance, and electronic support measures (ESM) systems.  To this end, in the following we discuss the direct detection of pulsed RF fields with an atomic receiver in the time-domain and investigate the behavior and response times of the atomic detector to both pulsed RF field detection and pulsed Rydberg EIT readout without RF to isolate the atom-optical interaction from the atom-RF interaction under typical EIT operating conditions.\\

Figure~\ref{fig2}(b) shows the time-domain detection and measurement of a 1~$\mu$s-long 36.2-GHz narrow-band RF field pulse by a rubidium Rydberg-based atomic detector.  The pulse-modulated RF pulse measurement is performed in the weak-field regime where the RF is resonant with rubidium 47S$_{1/2}$ to 47P$_{1/2}$ transition and AT-splits the EIT line following Eq.~1.  One observes in the measured data that the AT-splitting is well-resolved in time for 1~$\mu$s-long RF pulses.  The temporal resolution in the detection in Fig.~\ref{fig2}(b) approaches the $\sim 10$~ns level and is limited primarily by the response time of the photo-detector used in the measurements.  An extension to shorter RF pulse-width detection is readily achievable and corresponding larger RF detection bandwidths.  Development in this area is on-going.\\

In a closely related study of time-dependent effects, we have investigated the time-dependence of the underlying EIT readout from the atomic vapor for pulsed Rydberg EIT alone, without application of external RF fields.  This allows us to distinguish between atom-optical and atom-RF interaction effects contributing to the detection process, and to shed light on the short time-scale response of the Rydberg EIT pulse in a thermal atomic vapor for typical moderate optical Rabi frequencies. Figure~\ref{fig2}(c) shows the EIT probe transmission (gray-scale) for a 5~$\mu$s-long coupler-light pulse as a function of time (vertical axis) and coupler laser frequency (horizontal axis) near the field-free Rb 5P$_{3/2}$ to 30D$_{5/2}$ Rydberg state resonance.  Here, the coupler pulse is switched on at 11.7~$\mu$s and off again at 16.7~$\mu$s with a precision $<$100~ns.  When the coupler pulse turns on a sudden decrease in transmission is observed, or equivalently an increase in probe absorption, over a period of about 20~ns (white horizontal stripe in the data, labelled I in Fig.~\ref{fig2}(c)).  This is followed by an increase in transmission until reaching a steady-state value over a period of ~1 to 2~$\mu$s.  At the turn-off of the coupler pulse, a sudden increase (gain) in optical transmission is observed, also over about 20~ns (black horizontal stripe in the data, labelled II in Fig.~\ref{fig2}(c)), followed by a decay of the signal to zero over several microseconds.\\

The transients measured at both the beginning and the end of the Rydberg EIT coupler pulse in an atomic vapor have to our knowledge not been observed before.  The observed process appears akin to - but distinct from - photon storage and retrieval via EIT-mediated Rydberg polaritons in cold-atom systems, where a probe photon is stored as a collective Rydberg excitation in the medium in presence of the coupler beam and released/retrieved when the coupler is turned off~\cite{Dudin.2012,Peyronel.2012,Maxwell.2013}.  In our presented case, a certain excess amount of probe-pulse energy (contained in the probe light incident on the medium) is ‘stored’ and ‘released.’  In this interpretation, the 'stored' 780~nm light is 'retrieved' after an extremely long time ($>$10~$\mu$s-long pulses in other experiments), exceeding the $<$1~$\mu$s transit time of atoms through the EIT beams used.  The concept of collective Rydberg-polaritons propagating along the laser-beam direction, through a medium of atoms that are frozen in place (a picture commonly used in cold-atom EIT experiments~\cite{Dudin.2012,Peyronel.2012,Maxwell.2013}), is not directly applicable to our situation. However, during the short, $\sim 100$-ns-long time intervals that follow the light-switching events the atoms are approximately frozen in place, even in the presented case of Rydberg-EIT in a room-temperature vapor cell. This allows us the use of a frozen-atom model to explain the fast transients observed in Fig.~\ref{fig2}(c).\\

The transient responses of the Rydberg EIT readout studied here provide a time resolution at the sub-10~ns level.  Their implementation in RF field sensing is proposed here to achieve high-bandwidth receiving of modulated RF communications signals, short RF pulse detection, and high-frequency RF noise measurements.  In work not shown, we have modeled the Rydberg EIT system dynamics for the conditions in Fig.~\ref{fig2}(c) which provide Rydberg EIT transient dynamics that reproduce the observed transient behavior in great detail.  A detailed discussion and further characterization of the quantum physics and its implementation in ultra-fast RF detection method will be the subject of future work.\\

Comparing Figs.~\ref{fig2}(b) and~\ref{fig2}(c) it is noted that the EIT line widths are quite different. This is due to the use of different laser-beam parameters and Rydberg states, leading to different coupler- and probe-beam Rabi frequencies in the two cases. In Fig.~\ref{fig2}(b) the Rabi frequencies at the laser-beam centers are $\Omega_p = 2 \pi \times  18~$MHz for the probe and $\Omega_c = 2 \pi \times  2.5~$MHz for the coupler. These values are small enough to largely avoid saturation broadening, leading to  EIT lines that are less than about 10~MHz wide (in coupler laser frequency). In Fig.~\ref{fig2}(c), the respective Rabi frequencies are $\Omega_p = 2 \pi \times  44~$MHz and $\Omega_c = 2 \pi \times  10~$MHz. In that case, the large probe Rabi frequency causes a larger amount of saturation broadening, leading to EIT lines that are about 20~MHz wide.

\section{Atomic RF phase detectors}
\label{sec:phase}

RF electric-field sensing and measurement based on EIT readout of field-sensitive Rydberg states of atoms in thermal vapor cells has made rapid progress towards establishing atomic RF $E$-field standards.  Here we describe an atomic RF phase, amplitude, and polarization sensor that employs a novel quantum-optical readout scheme from an RF field-sensitive Rydberg vapor to achieve RF phase sensitivity~\cite{Anderson.2019}.\\

The holography concept outlined in Sec.~\ref{sec:background} can be translated from the optical into the RF domain. Measurements have been performed by combining RF signal and reference waves in or close to Rydberg-EIT vapor cells~\cite{Mingyong.2019,SimonsPhase.2019,GordonPhase.2019}. The magnitude of the coherent electric-field sum of the object and reference RF or microwave fields is measured using vapor-cell Rydbrg-EIT methods within the atomic vapor cell or hybrid atom-cavity cell structure, as described in Sec.~\ref{sec:field}.  According to principles of holography, this allows measurement of amplitude and phase of the signal wave, with the reference wave providing the phase reference as well as amplification~\cite{Mingyong.2019,SimonsPhase.2019,GordonPhase.2019}. Towards practical applications, a phase-sensitive recording of a coherent electromagnetic field on a surface allows the reconstruction of the field in all space.  RF-applications of this reconstruction principle are abound and include radars based on interferometric schemes, such as SAR and InSAR, and far-field characterization of antenna radiation patterns based on near-field measurements of amplitude and phase of the field emitted by the antenna under test. In the last application listed, the measurement has to be performed on a surface, and a near-field to far-field transformation is applied to calculate the field in all space.\\

To achieve phase sensitivity in the holographic RF field measurement, the reference wave can be interfered with the waves emitted by or reflected from an object. The generation of a clean RF reference wave presents a considerable problem. In optical holography, the reference wave typically is an expanded, near-perfect plane-wave laser beam that interferes with the object scatter within a layer of photographic emulsion (or an equivalent substance). The purity of the reference wave is important, {\sl{i.e.}} it should be free of diffraction rings caused by dust particles and other imperfections. Interference from specular reflections of the reference wave from planar surfaces should also be avoided. In quantitative work, it would also be important that the reference wave has a fixed amplitude or, at least, a well-known, slowly varying amplitude function. In holographic measurements in the RF domain, equivalent conditions are hard to meet. The preparation of a defect-free RF reference wave that has a smooth amplitude behavior over a large surface presents a great challenge. In some cases, it will be fundamentally impossible to prepare a stationary reference wave. This applies, for instance, to SAR radar applications, where the detector is mounted on a moving platform, like an airplane or a satellite, or in cases where a mm-wave or microwave field needs to be fully characterized over a large surface in space. In another class of applications, the object waves are located within close quarters where multiple reflecting surfaces cannot be covered with anechoic material (``urban radar''); there, reflections from unknown surfaces spoil the reference wave.\\

The cited previous implementations of holographic RF phase detection with atoms have required an antenna or similar for the generation of the reference RF wave, precluding the approach from providing a stand-alone atomic detector solution for RF waves propagating in free space~\cite{Mingyong.2019,SimonsPhase.2019,GordonPhase.2019}. In the following we present the holographic scheme in which an RF reference signal is provided via phase modulation of one of the EIT laser beams~\cite{Anderson.2019}. Our presented approach removes the need for RF reference waves, and therefore eliminates the aforementioned shortcomings of RF reference waves.\\

\begin{figure}[h]
\centerline{\includegraphics[width=\columnwidth]{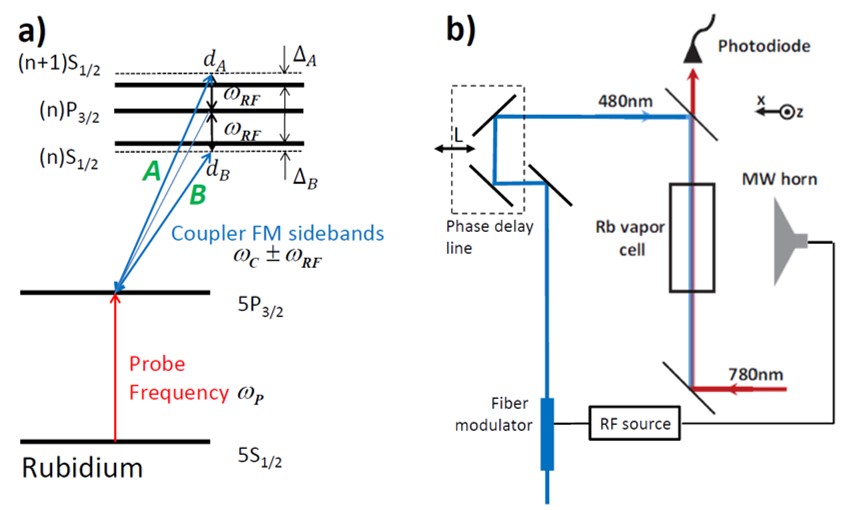}}
\caption{(a)  Quantum mechanical level scheme and optical/RF excitation pathways used in an implementation of the phase-sensitive RF electric-field measurement method~\cite{Anderson.2019}. (b) Setup illustration of the phase-sensitive measurement implementation. The microwave horn (MW) stands for any antenna under test or other object wave of interest. The fiber modulator phase-coherently imprints an RF reference beat onto the coupler beam sent to the atoms in the vapor cell. The RF reference beat replaces the reference beam that is normally used in phase sensitive (holographic) field measurement. The vapor cell in the atom-based RF sensing element can be very small ($\sim 1$~mm), fiber-coupled to the 780-nm and 480-nm laser beams. The phase of the optical reference beat is scanned via a mechanical optical delay line, as shown, or by an electro-optic control element.
}
\label{fig:phase1}
\end{figure}

For RF phase measurement using RF-modulated optical beams, we consider a phase modulation imprinted on an optical coupling laser beam via an electro-optic modulation technique. Using a fiber-optic high-frequency modulator, which is commercially available, the coupler beam is frequency- or amplitude-modulated with a signal at frequency $\omega_{RF}$ that is near the frequency of the RF field to be measured, and that is phase-coherent with the RF field to be measured.  For the purpose of describing the basic concept, in the following we consider a rubidium atom and a case where the (optical) coupler field is phase-modulated at a frequency that is identical with the RF signal frequency $\omega_{RF}$. Here, $\omega_{RF}$ also approximately equals half the separation between two neighboring Rydberg levels of rubidium, $nS_{1/2}$ and  $(n+1)S_{1/2}$, as shown in Fig.~\ref{fig:phase1}(a). The carrier frequency of the coupler laser beam is resonant with the forbidden transition $5P_{3/2}$ $\rightarrow$  $nP_{3/2}$. Due to the quantum defects in rubidium, the
$nP_{3/2}$ level is approximately at the midpoint between the $nS_{1/2}$ and  $(n+1)S_{1/2}$ levels, and the electric-dipole matrix elements for the allowed microwave transitions,  $d_{A}$ and $d_{B}$, are about the same. Also, the detunings of $\omega_{RF}$ from the respective atomic transition frequencies,
$\Delta_A$ and $\Delta_B$, are approximately equal in magnitude and opposite in sign (see Fig.~\ref{fig:phase1}(a)).
For the simplified discussion presented here, we assume that the detunings $\Delta_A$ and $\Delta_B$ have magnitudes on the order of 100~MHz, which is much larger than the Rabi frequencies of any of the involved transitions. Hence, the two-photon Rabi frequencies that describe the transitions from $5P_{3/2}$  into $nP_{3/2}$  via the absorption of one coupling-laser sideband photon and the absorption (channel B in Fig.~\ref{fig:phase1}(a)) or the stimulated emission (channel A in Fig.~\ref{fig:phase1}(a)) of an RF photon are given by
\begin{eqnarray}
 \Omega_A &=& \frac{\Omega_{5P(n+1)S} \, \Omega_{(n+1)SnP}}{2 \Delta_A}
              \exp( {\rm{i}}(\phi_{5P(n+1)S}-\phi_{RF}))  \nonumber \\
 \Omega_B &=& \frac{\Omega_{5PnS} \, \Omega_{nSnP}}{ 2 \Delta_B}
               \exp( {\rm{i}}(\phi_{5PnS}+\phi_{RF}))
\end{eqnarray}
There, $\Omega_{5PnS}$ and $\Omega_{5P(n+1)S}$ are the Rabi frequencies of the optical coupler-laser transitions into the $S$ Rydberg levels, $\Omega_{nSnP}$ and $\Omega_{(n+1)SnP}$ are the Rabi frequencies of the RF transitions from the $S$ Rydberg levels into the $nP_{3/2}$ Rydberg level, and $\phi_{RF}$ is the phase of the RF field.  Also, $\phi_{5PnS}$ and $\phi_{5P(n+1)S}$  are the phases of the modulation sidebands of the coupling laser.  Note there is an important difference in sign in front of the $\phi_{RF}$ in the above equations. Further, the RF field amplitude, $E$, is included in the RF Rabi frequencies. It is, for instance, $\Omega_{5PnS} = E d_B / \hbar$. The net coupling, $\Omega_C$, due to the coupler lasers is given by the coherent sum of $\Omega_A$ and $\Omega_B$.  Noting that $\Omega_{5PnS} \approx$ $\Omega_{5P(n+1)S}$ and $\Omega_{nSnP} \approx$ $\Omega_{(n+1)SnP}$, and noting that a suitable choice of levels allows us to set $\Delta_B = - \Delta_A =: \Delta$, for the present simplified discussion we have

\begin{equation}
\Omega_C = \frac{\Omega_{5PnS}\, \Omega_{nSnP}}{2 \Delta} (\exp( {\rm{i}}(\phi_{5P(n+1)S}-\phi_{RF}))
- \exp( {\rm{i}}(\phi_{5PnS}+\phi_{RF})))
\end{equation}

The approximations made to arrive at this expression are not crucial; they serve to simplify the math to better elucidate the important aspects of the method. The optical phases  $\phi_{5PnS}$ and $\phi_{5P(n+1)S}$  are well-defined and are not prone to drift, because all frequency components of the modulated coupling laser beam follow the exact same geometrical path. An optical delay line in the beam path of the FM-modulated coupler laser is used to control the difference between the optical phases $\phi_{5PnS}$ and $\phi_{5P(n+1)S}$. A translation by amount $L$ (see  Fig.~\ref{fig:phase1}(b)) causes a phase shift of $\frac{4L}{c} \omega_{RF}$. For RF frequencies in the 10-GHz range, a translation of about 1~cm will scan the optical-phase difference $\phi_{5PnS} - \phi_{5P(n+1)S}$ over a range of $2 \pi$. It is seen from the previous equation that the net EIT coupling takes the form
\begin{equation}
\Omega_c = \Omega_{c0} \cos(\phi_{RF} + \phi_{opt}) \quad ,
\end{equation}
with a (complex) factor $\Omega_{c0}$ that neither depends on $\phi_{RF}$ nor on the delay-line-controlled
optical phase $\phi_{opt}$. It is thus seen that net EIT coupling Rabi frequency $\Omega_c$ can be tuned between zero
and  $\pm \Omega_{c0}$ by adjusting the optical phase $\phi_{opt}$ with the coupler-beam delay line (see  Fig.~\ref{fig:phase1}(b)). The presented analysis shows that the optical phase $\phi_{opt}$ is equivalent with the tunable offset phase $\phi_{ofs}$ in the introductory discussion~\ref{sec:background}. Also, $\phi_{RF}$ corresponds with the frequency difference $\phi - \phi{ref}$ that is to be measured.\\

Since the strengths of the Rydberg-EIT lines observed in the spectra are generally proportional to $\vert \Omega_{c} \vert^2$, the EIT line strength is proportional to $\cos^2(\phi_{RF} + \phi_{opt})$. The EIT line strength, measured
as a function of the optical phase, $\phi_{opt}$, allows one to measure the phase
$\phi_{RF}$. The microwave phase $\phi_{RF}$ can therefore be retrieved as long as it remains stable over
the time scale needed to scan the optical delay line over a range of $2\pi$. Using mechanical delay lines, the dynamic range of this RF phase measurement method will be at about 10~s$^{-1}$, while electro-optic phase shifters will allow a phase measurement bandwidth ranging into the MHz-range.\\

We note that in the presented scheme the $5P_{3/2}$ to $nP_{3/2}$ transition is forbidden; therefore, the coupler-beam carrier (thin blue line in Fig.~\ref{fig:phase1}(a)) does not introduce an additional term in the analysis. In more general cases, such a term could, of course, be included. Further, the magnitude of the pre-factor $\Omega_{c0}$ can be determined by finding the peak EIT line strength while varying $\phi_{opt}$. The obtained peak value for $\Omega_{c0}$ then reveals the RF electric field, $E$, via the known electric dipole moments of the RF transitions. In this way, both
$E$ and the phase $\phi_{RF}$ can be measured. This capability enables the aforementioned applications in antenna characterization, phase-sensitive radar, communications, and sensing.

\section{Conclusion}

In this work we have demonstrated the capability of pulsed RF field detection and measurement with an atomic receiver.  Pulsed RF field detection was performed in the time-domain with a temporal resolution at the 10~ns-level, limited by photodetector bandwidths. The behavior and response times of the atomic detector to both pulsed RF field detection and pulsed Rydberg EIT readout without RF have been investigated to isolate the atom-optical interaction from the atom-RF interaction under typical EIT operating conditions.  In pulsed Rydberg EIT readout from the atomic vapor, transient behavior was experimentally observed resulting in a drop in optical transmission at the onset of the coupler pulse and gain in optical transmission at the turn-off of the coupler pulse with dynamics on a 10~ns timescale, also limited by photodetector bandwidth.  Modeling of these system dynamics has separately been performed reproducing the observed transient behavior in great detail and affirming the physical existence of the phenomenon, with underlying physics distinct from the interpretation of collective Rydberg-excitation polaritons propagating in the medium~\cite{Dudin.2012,Peyronel.2012,Maxwell.2013}.  Fast quantum-optical transient dynamics in Rydberg EIT readout at time-scales on the sub-10~ns level have been studied, and their implementation in RF field sensing has been proposed to enable, for example, reception of modulated RF communications signals approaching 100~MHz bandwidth, short RF pulse detection, and high-frequency RF noise measurements.  In the present work, we also describe a new method for atomic RF phase sensing and measurement to realize atomic sensors for phase-sensitive detection of RF fields~\cite{Anderson.2019} critical to a wide range of application areas such as antenna near-field characterizations, radar based on interferometric schemes, and phase-modulated signal transmission and telecommunications.  The atomic RF phase sensor development enables the realization of atomic sensors, receivers and measurement tools capable of RF phase, amplitude, and polarization detection with a single, vapor-cell sensing element.  Atomic RF sensors and receivers based on Rydberg atom-mediated RF-to-optical transduction hold promise as a basic technology platform to realize advanced passive radar and electronic support measures (ESM) systems.  Implementation of coherent conversion between microwave optical photons in Rydberg gases~\cite{Han.2018}, for example, may be implemented in the Rydberg atom-based detector platform to realize coherent RF-to-optical transducers in quantum communications schemes and radar.

\section{Acknowledgement}
This work was supported by Rydberg Technologies Inc.  Part of the presented material is based upon work supported by the Defense Advanced Research Projects Agency (DARPA) and the Army Contracting Command-Aberdeen Proving Grounds (ACC-APG) under Contract Number W911NF-17-C-0007.  The views, opinions and/or findings expressed are those of the author and should not be interpreted as representing the official views or policies of the Department of Defense or the U.S. Government.




\end{document}